# Evaluation of Link Traversal Query Execution over Decentralized Environments with Structural Assumptions

Ruben Taelman

Ruben Verborgh

IDLab, Department of Electronics and Information Systems, Ghent University – imec
Ghent, Belgium
E-mail: ruben.taelman@ugent.be

## ABSTRACT

To counter societal and economic problems caused by data silos on the Web, efforts such as Solid strive to reclaim private data by storing it in permissioned documents over a large number of personal vaults across the Web. Building applications on top of such a decentralized Knowledge Graph involves significant technical challenges: centralized aggregation prior to query processing is excluded for legal reasons, and current federated querying techniques cannot handle this large scale of distribution at the expected performance. We propose an extension to Link Traversal Query Processing (LTQP) that incorporates structural properties within decentralized environments to tackle their unprecedented scale. In this article, we analyze the structural properties of the Solid decentralization ecosystem that are relevant for query execution, and provide the *SolidBench* benchmark to simulate Solid environments representatively. We introduce novel LTQP algorithms leveraging these structural properties, and evaluate their effectiveness. Our experiments indicate that these new algorithms obtain accurate results in the order of seconds for non-complex queries, which existing algorithms cannot achieve. Furthermore, we discuss limitations with respect to more complex queries. This work reveals that a traversal-based querying method using structural assumptions can be effective for large-scale decentralization, but that advances are needed in the area of query planning for LTQP to handle more complex queries. These insights open the door to query-driven decentralized applications, in which declarative queries shield developers from the inherent complexity of a decentralized landscape.

## 1 INTRODUCTION

Despite transforming our world to be more interconnected than ever before, the Web has become increasingly *centralized* in recent years, contrary to its original vision [1]. Today, the majority of data on the Web is flowing towards isolated *data silos*, which are in the hands of large companies. This siloization of data leads to various problems, ranging from issues with cross-silo interoperability and vendor lock-in, to privacy issues and the fact that individuals' data is controlled by companies instead of themselves.

Because of these reasons, and the introduction of user-empowering legislature such as GDPR and CCPA, decentralization initiatives [2,3,4] are gaining popularity. Their common goal is to give people back control over their own data by guarding it in chosen locations on the Web instead of aggregated in silos. Initiatives such as Solid [2] do this by allowing users to store any kind of data in their own personal data vault, which they fully control. These data vaults form personal Knowledge Graphs [5], which are often represented as collections of Linked Data documents [6] containing RDF triples [7]. The presence of such data vaults results in a large-scale distribution of data, where applications involving multiple individuals' data require accessing thousands or even millions of documents across different data vaults across the Web. These applications cannot effectively be built today due to the lack of querying techniques that can hande the requirements of decentralized environments like Solid.

The majority of research in the domain of query execution over Knowledge Graphs on the Web has been focused on centralized use cases, where all data is captured in a single or a small number of sources, usually exposed as SPARQL endpoints [8]. Even though several federated query execution approaches exist [9,10,11,12], they have been designed for federating over *a few (~10) large sources*, while decentralized environments such as Solid are identified by *a large number (~millions) of small sources*. Furthermore, many federated query execution techniques assume all sources to be known prior to query execution, which is not feasible in decentralized environments due to the lack of a central index. Hence, these techniques are unsuitable for decentralized environments.

*Link Traversal Query Processing (LTQP)* [13,14] is an alternative query execution paradigm that is more promising for uncharted decentralized environments. It can query over a continuously growing range of documents that are discovered during query execution, by *following hyperlinks* between Linked Data documents using the *follow-your-nose* principle [6]. While LTQP has mainly been a theoretically interesting technique, it has not seen any practical use so far, in particular because of performance concerns. While the Linked Data principles [6] provide us with the ability to apply the follow-your-nose principle, decentralized environments such as Solid provide additional structural properties on top of these principles that allow us to make additional assumptions about data, documents, and their organization during query execution. For example, Solid makes use of the Linked Data Platform specification [15] to provide completeness guarantees when finding data within vaults, and it provides the Type Index [16] to enable type-based document discovery.

In this work, we prove that LTQP can be an effective paradigm, if we exploit specific structural properties within decentralized environments for more effective source discovery and query optimization. We apply our research to the Solid ecosystem, but these concepts may be generalizable to other decentralization initiatives [3,4]. To the best of our knowledge, this is the first in-depth analysis of query execution within the Solid ecosystem.

This article is structured as follows. In the next section, we discuss the related work, after which we provide an analysis of the structural properties of Solid data vaults in Section 3. Next, in Section 4 we provide a benchmark that simulates a decentralized Solid environment based on this analysis. In Section 5, we introduce LTQP algorithms that make use of these structural properties, which are evaluated in Section 6. Finally, we conclude in Section 7.

Specifically, we provide the following contributions in this work:

- An analysis of the Solid ecosystem from a query execution perspective.
- SolidBench: A benchmark that simulates a decentralized Solid environment, with dedicated fragmentation techniques, choke-point-based query templates, and LTQP-specific metrics.
- Two novel and formally-defined LTQP discovery algorithms over the structural properties in the Solid ecosystem.
- Implementation of these novel algorithms and the existing foundational LTQP algorithms.
- Extensive evaluation of the implemented algorithms using SolidBench.

## 2 RELATED WORK

Below, we discuss related work on Knowledge Graphs, Link Traversal Query Processing, and related benchmarks.

### 2.1 Knowledge Graphs On The Web

The standard for modeling knowledge graphs is the Resource Description Framework (RDF) [7]. Fundamentally, it is based around the concept of *triples* that are used to make statements about *things*. A triple is made up of a *subject*, *predicate* and *object*, where the *subject* and *object* are resources (or *things*), and the *predicate* denotes their relationship. Resources can either be identified by *Uniform Resource Identifiers (URIs)*, or by blank nodes. Multiple resources can be combined with each other through multiple triples, which forms a *Knowledge Graph*. The Linked Data principles [6] is a set of best-practises for publishing RDF on the Web, such as using URIs to identify resources, and returning RDF using a standard serialization like Turtle [17] and JSON-LD [18] when looking up those URIs. In order to look up information within Knowledge Graphs, the SPARQL query language [19] was introduced as a standard. Essentially, SPARQL allows RDF data to be looked up through combinations of *triple patterns*, which can make up arbitrarily complex queries.

### 2.2 Link Traversal Query Processing

The Link Traversal Query Processing (LTQP) paradigm was introduced more than a decade ago [20] as a way to query over the Web of Linked Data as if it was a globally distributed dataspace, without having to first index it in a single location. LTQP does this by employing the *follow-your-nose* principle of Linked Data [6] during query execution, where new RDF are continuously added to a local dataset while discovering new sources by following links between documents. An iterator-based pipeline [20] allows query execution to take place without having to wait until all links have been followed. As LTQP engines may traverse over documents from untrusted publishers, specific security vulnerabilities [21] arise that do not apply when querying in a controlled environment.

LTQP is thus an *integrated* approach [22] with parallel source discovery and query execution, in contrast to *two-phase* approaches [23,24] that perform data retrieval and indexing *before* query execution. As a one-phase approach, LTQP cannot rely on traditional pre-execution optimization algorithms that require prior dataset statistics. A zero-knowledge query planning technique [25] instead orders triple patterns in a query based on link traversal-specific heuristics. LTQP is related to the idea of *SQL-based query execution over the Web* [26,27] and to the concept of *focused crawling* [28,29]. While LTQP considers the Web of Linked Data a large database using the RDF data model, SQL-based approaches focus on querying attributes or content within Web pages. Focused crawlers search for Web pages of specific topics to populate a local database or index, using a two-phase approach where a preprocessing step precedes execution. While two-phase approaches in general are able to produce better query plans using traditional cardinality-based planning techniques, waiting for the data retrieval phase to be completed may be impractical or even impossible for certain queries.

The number of potential links to be followed within the Web of Linked Data can become prohibitively large. In the worst case, a single query could theoretically require traversing the entire Web. Therefore, the formal LTQP model [14] enables different *reachability criteria*, which embody strategies for deciding what links to follow, each leading to different result completeness semantics. *cNone* follows no URLs, *cAll* follows all URLs in all encountered triple components, and *cMatch* only follows URLs in triple components for those triples that match a triple pattern within the query. *Context-based semantics* [30] is an extension of these reachability semantics to cope with property path expressions in the SPARQL 1.1 language [19]. Next to query-driven reachability, another extension [31] introduces the ability for data publishers to express which links should be followed using *subweb specifications*.

Next to filtering links via different semantics, a second methodology for improving query result arrival times is through *link prioritization* [32]. However, existing techniques only sometimes result in faster query results compared to no prioritization.

Even though multiple query languages [33,34,35] have been introduced specifically for LTQP, its SPARQL-based execution model [20] is still the most widely used. Since SPARQL is the only language among these that is a standard, and the fact that it is more widely known and supported by different tools, we make use of it within this work. Nevertheless, the concepts within this work can be applied to other languages as well.

In general, LTQP has mostly been applied to querying Linked Open Data on the Web. This is in contrast to our work, where we apply LTQP to specific decentralized environments, where specific structural properties apply, and not all data is open and accessible to everyone. Privacy issues [36] that arise because of the private data in decentralized environments are considered out of scope for this work.

### 2.3 Link Traversal Benchmarks

Aside from the lack of reusable link traversal query engines, there is also a lack of easily reusable benchmarks for evaluation the performance of link traversal query engines. According to a recent survey of Linked Data querying approaches [22], *"there are no well-defined and well-understood benchmarks to test Linked Data query processing systems"*. Nevertheless, there has been work that provide a foundation onto which benchmarks may be built.

QWalk [37] is a methodology for building a large set of queries based on random walks through an existing dataset. The output of this methodology is a set of Basic Graph Patterns that crosses through various documents via dereferenceable links. QWalk has been designed to be executed on the actual Web, to evaluate uncontrolled environments. Due this this, most of the queries generated by the authors make use of links that have gone dead by now, which makes these queries unsuitable for a reliable and reusable benchmark. Furthermore, the authors propose using queries from existing benchmarks such as FedBench [38] and the DBpedia SPARQL Benchmark [39] for execution over the actual Web. Unfortunately, these also lead to inconsistent results, which makes them unsuitable for reliable benchmarking. In contrast, our benchmark provides a closed environment in full control of the experimenter.

SPLODGE [40] is similar to QWalk, and makes use of crawled Linked Open Data. It therefore also suffers from the issues arising from uncontrolled execution environments like QWalk. Additionally, since it relies on crawled public data, it is not able to capture the structural properties specific to decentralized environments such as Solid, as these may be hidden behind an authentication layer. Our benchmark does not rely on crawled data, and instead simulates these structural properties.

WODSim [32] is a tool that accepts an RDF dataset as input, and is able to simulate a Web of Linked Data documents. For each triple in the dataset, it can either place the triple inside the Linked Data document(s) identified by the triple's subject, object, or both. While this tool is beneficial for running experiments in a reproducible manner, it does not ship with a standard dataset, which makes is less convenient for reusability. Our benchmark builds on top of the WODSim approach for fragmenting triples into documents, but also provides a standard dataset.

Aside from these approaches, evaluation of link traversal engines is usually done via hand crafted queries [20,41,25,42].

## 3 THE SOLID ECOSYSTEM

In this section, we provide an analysis of the structural properties within the Solid ecosystem that are relevant for query processing. We start by explaining the concept of data vaults and its implications on applications. Next, we explain the WebID, which is used for identifying users. Then, we discuss the Solid type index; a structural property that improves data discovery. Finally, we list requirements for query processing within the Solid ecosystem.

### 3.1 Data Vault

A primary element within the Solid Protocol [43] is the data vault (also known as *data pod*), which is a user-controlled space in which any kind of data can be stored. Users can choose where and how their vault is stored on the Web, by hosting it themselves [44], obtaining service-provided space by a company [45] or government [46]. Data vaults are intended to be loosely coupled to applications, and applications must request explicit access to the user for interacting with specific data. This loose coupling enables different applications to use the same data interoperably.

Current data vaults are primarily document-oriented, and are exposed on the Web as a REST API using elements of the Linked Data Platform (LDP) specification [15]. Directories are represented using *LDP Basic Containers*, which can contain any number of resources that correspond to RDF or non-RDF documents (via `ldp:contains` links), or other nested basic containers. For the remainder of this article, we will only consider the processing of RDF documents within vaults. Resources within vaults can be read by sending HTTP GET requests to their URLs, with optional content negotiation to return the documents in different RDF serializations. If the vault supports this, resources can be modified or created using HTTP PATCH and POST requests. An example of such a basic container can be found in the Listing 1.

```
@prefix ldp: <http://www.w3.org/ns/ldp#>.
<> a ldp:Container, ldp:BasicContainer, ldp:Resource;
  ldp:contains <file.ttl>, <posts/>, <profile/>.
<file.ttl> a ldp:Resource.
<posts/> a ldp:Container, ldp:BasicContainer, ldp:Resource.
<profile/> a ldp:Container, ldp:BasicContainer, ldp:Resource.
```

**Listing 1: An LDP container in a Solid data vault containing one file and two directories in the Turtle serialization.**

Data vaults can contain public as well as private data. Users can configure who can access or modify files within their vault using mechanisms such as ACL [47] and ACP [48]. This configuration is usually done by referring to the *WebID* of users.

### 3.2 WebID Profile

Any agent (person or organization) within the Solid ecosystem can establish their identity through a URI, called a *WebID*. These agents can authenticate themselves using the decentralized Solid OIDC protocol [49], which is required for authorizing access during the reading and writing of resources.

According to the WebID profile specification [50], each WebID URI should be dereferenceable, and return a WebID profile document. Next to basic information of the agent such as its name and contact details, this document should contain links to 1) the root LDP container of its data vault (via `pim:storage`), and 2) public and private type indexes. An example is shown in Listing 2. We omit further WebID details due to their irrelevance within this work.

```
@prefix pim: <http://www.w3.org/ns/pim/space#>.
@prefix foaf: <http://xmlns.com/foaf/0.1/>.
@prefix solid: <http://www.w3.org/ns/solid/terms#> .
<#me> foaf:name "Zulma";
    pim:storage </>;
    solid:oidcIssuer <https://solidcommunity.net/>;
    solid:publicTypeIndex </publicTypeIndex.ttl>.
```

**Listing 2: A simplified WebID profile in Turtle.**

### 3.3 Type Index

The Type Index [16] is a document that enables type-based resource discovery within a vault. Users may have public or private type indexes, which respectively refer to data that are and are not publicly discoverable. A type index can contain type registration entries for different classes, where each registration has a link to resources containing instances of the corresponding class. Listing 3 shows a type index example with type registrations for posts and comments, where the posts entry refers to a single posts file, and the comments entry refers to a container with multiple comments files. If an application wants to obtain all posts of a user, it can do so by finding this type index and following the link within the type index entry that corresponds to the post class.

```
@prefix ldp: <http://www.w3.org/ns/ldp#>.
<> a solid:TypeIndex ;
    a solid:ListedDocument.
<#ab09fd> a solid:TypeRegistration;
    solid:forClass <http://example.org/Post>;
    solid:instance </public/posts.ttl>.
<#bq1r5e> a solid:TypeRegistration;
    solid:forClass <http://example.org/Comment>;
    solid:instanceContainer </public/comments/>.
```

**Listing 3: Example of a type index with entries for posts and comments in the Turtle serialization.**

### 3.4 Requirements For Query Engines

Instead of requiring application developers to reinvent the wheel by manually discovering application-relevant elements within data vaults, discovery can be abstracted away behind declarative queries. This makes applications robust against changes or additions within the Solid Protocol. The application's declarative query can remain unchanged, whereas the underlying query engine can be updated to improve the result set.

The Solid protocol [43] only establishes a minimal set of groundrules to make data vaults and applications interoperable. Below, we list additional query agent requirements for enabling querydriven Solid applications over data vaults with a sufficient level of user-perceived performance [51].

1. **Mapping query to a sequence of HTTP requests**: Convert a query into a sequence of HTTP requests across data vaults.

2. **Discovery and usage of LDP storage**: Discover and follow the link in a WebID profile to the storage root of the vault, identify an LDP basic container, and follow (a subset of) links towards the resources within this container.

3. **Discovery and usage of type indexes**: Discover and follow type index links from the WebID profile, and handle (a subset of) the type registration links.

4. **Variability of vault structures**: Make no assumptions about the location of certain data within vaults without an explicit and discoverable link path to it, e.g. via LDP storage or type indexes. This is important for the interoperability of Solid apps because different apps or user preferences may lead to the storage of similar data in different locations within vaults.

5. **Authenticated requests**: Perform authenticated HTTP requests on behalf of the user after authentication with a WebID, to enable querying over private resources.

While these requirements apply to both read and write queries, we will focus on read-only queries for the remainder of this article. Furthermore, given the Linked Data nature of Solid, we will focus on queries using the SPARQL query language [19] for the remainder of this article. Nevertheless, these concepts can also be applied to write-queries and other query languages.

## 4 BENCHMARK

In this section, we introduce *SolidBench*, a benchmark that enables reproducible performance measurements of different query execution approaches within a decentralized environment. As discussed in Section 2, there is a need for this as existing benchmarks 1) do not provide a closed and reliable Web environment, 2) do not ship with standard datasets, and 3) are not expressive enough to capture the structural properties in the Solid ecosystem. SolidBench simulates a decentralized Solid environment with corresponding workload representative of a social networking application. Hereafter, we start by explaining the design considerations of the benchmark and our use case scenario, after which we introduce an overview of SolidBench. Next, we zoom in on important details of the benchmark, such as fragmentation of the data and the query workload.

## 4.1 Design Considerations

The goal of our benchmark is to simulate a realistic decentralized environment based on the Solid ecosystem, and provide a realistic workload that simulates a Solid application that reads data from one or more vaults. To develop this benchmark, we built upon the choke point-based design methodology from the Linked Data Benchmark Council [52].

Based on our analysis of the Solid ecosystem in Section 3, and requirements from similar benchmarks [52,53,54], we introduce the following requirements for our benchmark:

1. **WebIDs**: Dataset consists of WebIDs corresponding to simulated agents. Each WebID refers to one LDP-based storage vault, and is represented using a standard RDF serialization.
2. **Data vaults**: Dataset consists of data vaults containing RDF files in LDP containers.
3. **Type indexes**: WebIDs link to type indexes, containing registrations for data in the agent's vault.
4. **Variability of vaults**: Data is organized differently in across vaults, to simulate different query operations and data organization preferences.
5. **Scalable dataset**: Dataset is configurable in the number of vaults and vault sizes.
6. **Workload**: Queries that evaluate different query operations and data linking structures.
7. **Metrics**: Measuring metrics such as total query execution time, result arrival times, number of HTTP requests, correctness, and completeness.
8. **Configuration**: Configurable in terms of queries and dataset properties, with default values.

## 4.2 Social Network Scenario

Since the original goal of the Solid project was to enable social interactions in a decentralized manner, the use case scenario our benchmark tackles is that of a social network. Concretely, different users each have their own data vault, and each user can provide personal details about themselves in their WebID document such as name, city of residence, and birthday. Next to that, users can express unidirectional knowledge relationships to other users. Furthermore, users can create posts, or leave comments on other posts. To stay in line with the ownership model of Solid, each post or comment a user creates, is stored within that user's data vault. Hence, chains of comments can span multiple data vaults.

## 4.3 SolidBench Overview

We build upon the well-established Social Network Benchmark (SNB) [54,55], which models a social network akin to Facebook, and meets most of the requirements of our desired use case scenario. Since SNB was designed to evaluate the performance of centralized query engine, its dataset generator outputs a dataset in a single large dataset. Given that we aim to simulate a decentralized social network, we introduce a *fragmentation layer* on top of this generator. This fragmenter is able to take in any dataset as input, and provide a fragmented version of this dataset that simulates an interlinked set of Linked Data documents, inspired by WODSim [32].

We introduce the following tools with SolidBench:

- **Dataset generator**: consisting of SNB's existing generator, and a new dataset fragmenter.
- **Query generator**: consisting of SNB's existing generator, and a new fragmentation-aware query template instantiator.
- **Validation generator**: building on top of SNB's validator, produces fragmentation-aware validation sets containing queries and expected results.
- **Dataset server**: serving of fragmented datasets over HTTP with content negotiation for all RDF serializations.
- **Benchmark runner**: incorporation into an existing benchmarking system for execution against query engines via the SPARQL protocol [8].

For the query workload, we build upon the *interactive* workload of SNB, and extend it with additional queries to cover link-related choke points. Since the query templates that are produced by the generator of SNB assume a centralized dataset, we also add a layer on top of these query templates that can transform the queries to correspond to the decentralized dataset. We provide 27 query templates that can be instantiated any number of times to simulate a query workload. We also provide a tool that can produce validation queries and results to measure the correctness and completeness of a system. Since we focus on read-only queries in this work, we do not consider the write queries of SNB.

By default, SolidBench sets the *scale factor* of the SNB generator to 0.1, which results in 158.233 RDF files over 1.531 data vaults using the default fragmentation strategy. In total, there are 3.556.159 triples across all files, with an average of 22,47 triples per file. This scale primarily determines the number of persons in the dataset, which directly corresponds to the number of data vaults that will be simulated. Even though this scale can be increased arbitrarily, we notice that this default scale can already stress existing LTQP approaches beyond their current capabilities. Next to this vault scale factor, we also provide a new *multiplication factor* for the amount of posts inside a vault, which allows increasing vault sizes to arbitrary amounts. By default, this post multiplication factor is set at 1. When setting this to 5, the total number of triples is 9.404.520 (average of 29.75 triples per file, across 316.053 files). For more details on properties of this dataset and its schema, we refer to the SNB papers [54,55].

All aspects of SolidBench are fully configurable using JSON-LD configuration files [56], ranging from fragmentation strategies to properties of query templates. Furthermore, our benchmark is included in the benchmark runner *jbr* (https://github.com/rubensworks/jbr.js), which simplifies its execution. To simplify evaluation and testing, we also provide a built-in Web server that can serve the generated data vaults over HTTP using a single command, which is done using a slimmed-down version of the Community Solid Server [44]. This server disables authentication and authorization by default, so experiments can focus on query performance. The benchmark is open-source at https://github.com/SolidBench/SolidBench.js.

## 4.4 Fragmentation

To convert the centralized dataset produced by SNB into a decentralized environment, we provide a tool to fragment datasets using different fragmentation strategies. While this tool is highly configurable in terms of its strategies using declarative JSON-LD-based configuration files, we summarize its functionality in terms of two fragmentation dimensions for brevity. Finally, we illustrate this functionality by discussing fragmentation strategies of posts within the SNB dataset. All strategies within this tool are implemented in a streaming manner, which means that it can handle input datasets of any size, and the dataset does not have to be loaded fully in memory before it can be processed.

**Fragmentation dimensions**

*Triple document assignment* is the first dimension of fragmentation, which concerns the task of deciding which triples are placed in what files. Inspired by WODSim [32], we provide subject and object-based approaches, which respectively place each triple in the file referred to by their subject or object. These approaches can also be combined to place triples in both files referred to by subject and object. Additionally, we provide composition-based approaches, using which triples matching certain triple patterns can be assigned to a different approach. The second dimension of fragmentation is that of *URI rewriting*, in which URIs can be modified to eventually end up in different documents according to the first dimension. For example, this allows URIs matching a regex to be modified, or triples to be appended upon matching a triple pattern.

**Strategies for fragmenting posts and comments**

Based on the two fragmentation dimensions discussed above, our fragmenter can be configured to manage different fragmentation strategies for posts and comments in the SNB data schema. While these strategies can be applied to both posts and comments, we only explain them hereafter in terms of posts:

1. **Separate**: Each post created by a person is placed in a separate RDF file within that person's vault.

2. **Single**: Posts created by a person are placed in a single RDF file within that person's vault.

3. **Location-based**: Posts created by a person are placed in files in that person's vault corresponding to the location at which the post was created.

4. **Time-based**: Posts created by a person are placed in files in that person's vault corresponding to the post creation day.

5. **Composite**: The strategies above are assigned randomly to all persons in the dataset.

By default, SolidBench makes use of the composite strategy, which results in fragmentation variance across the different vaults, which is realistic for the Solid ecosystem.

## 4.5 Workload

As mentioned above, we make use of the *interactive* workload of SNB, since these correspond to the workload that social network applications would produce. We consider other SNB workloads (such as the business intelligence workload) out of scope, since these perform dataset analytics, which requires access to the whole dataset, which is not feasible in decentralization environments such as Solid where data can reside behind access control. Furthermore, we also focus solely on the class of read-only queries due to the scope of this article, but our approach can be extended towards write queries.

The SNB interactive workload consists of two classes of query templates: *short* and *complex* read queries. Since these queries cover the choke points related to linking structures only partially, we add *discover* query templates as third class.

**Choke points**

Following the choke point-based design methodology [52], the short and complex SNB interactive workloads already cover the majority of the 33 choke points introduced by SNB. We refer to the SNB specification [55] for more details on the correlation of short and complex query templates to these choke points. Since the short and complex query classes only partially cover choke points related *linking structures* within data vaults, we introduce *discover* queries dedicated for covering these choke points on linking structures.

The additional choke points related to *linking structures* within data vaults we introduce are the following:

‣ **CP L.1: Traversal of 1 hop (1)**: Following one link to one other document.

‣ **CP L.2: Traversal of 1 hop (n)**: Following one link to multiple other document.

‣ **CP L.3: Traversal of 2 hops (1:1)**: Following one link to another document, and one link to another documents.

‣ **CP L.4: Traversal of 2 hops (1:n)**: Following one link to another document, and multiple links to other documents.

‣ **CP L.5: Traversal of 3 hops (n:1:1)**: Following multiple links to other documents, one link from the next document, and one other link.

‣ **CP L.6: Traversal of 3 hops (n:1:n)**: Following multiple links to other documents, one link from the next document, and multiple other link.

‣ **CP L.7: Fragmentation variability in vaults**: Handling the variability of data fragmentation across different data vaults.

‣ **CP L.8: Index delegation**: The potential of deferring subqueries to an index (such as the type index).

‣ **CP L.9: Noise**: The ability to filter out HTTP requests that are irrelevant to the query.

The discover queries we introduce are listed below:

‣ D1: All posts of a person

- D2: All messages of a person
- D3: Top tags in messages from a person
- D4: Top locations in comments from a person
- D5: All IPs a person has messaged from
- D6: All fora a person messaged on
- D7: All moderators in fora a person messaged on
- D8: Other messages created by people that a person likes messages from

The correlation of these choke points to the discover queries is summarized in Table 1.

| Choke Point | D1 | D2 | D3 | D4 | D5 | D6 | D7 | D8 |
|---|---|---|---|---|---|---|---|---|
| L.1 | ✓ | ✓ | | | | | | |
| L.2 | | | | | ✓ | | | |
| L.3 | | | | | | ✓ | | |
| L.4 | | | ✓ | ✓ | | | | |
| L.5 | | | | | | | ✓ | |
| L.6 | | | | | | | | ✓ |
| L.7 | ✓ | ✓ | ✓ | ✓ | ✓ | ✓ | ✓ | ✓ |
| L.8 | ✓ | ✓ | | ✓ | | | | ✓ |
| L.9 | ✓ | ✓ | ✓ | ✓ | ✓ | ✓ | ✓ | ✓ |

**Table 1: Coverage of choke points on linking structures for discover queries.**

More details on all query templates can be found at https://github.com/SolidBench/SolidBench.js/blob/master/templates/queries/README.md.

**Query template instantiation**

Our benchmark contains 27 query templates, from which 19 are derived from queries within SNB. These query templates can be instantiated multiple times for different resources, based on the dataset scale. By default, each template is instantiated five times, so that metrics can be averaged to reduce the effect of outliers. Due to the fragmentation and URI rewriting we apply on top of the SNB dataset, we were unable to make use of the standard SNB query templates and its method of query instantiation. Therefore, we have implemented a custom query template instantiation tool that takes into account these fragmentations.

An example of discover query 8 can be found in Listing 4, which covers the majority of choke points related to linking structures. It is instantiated with a person's WebID URI, and finds all messages created by people that this person likes messages from. Since it starts from all liked messages of the starting person, then navigates to the creator of those messages, and then retrieves the contents of those messages, it covers CP L.6. Furthermore, since messages are fragmented in different ways across vaults, and the query spans different vaults, it covers CP L.7. Since messages can be captured within the user's type index, CP L.8 is also covered. Finally, since only message-related documents needs to be retrieved from the vaults, all other documents could potentially pruned out, covering CP L.9.

```
PREFIX snvoc: <http://localhost:3000/www.ldbc.eu/ldbc_socialnet/1.0/vocabulary/>
SELECT DISTINCT ?creator ?messageContent WHERE {
    ?person snvoc:likes [ snvoc:hasPost|snvoc:hasComment ?message ].
    ?message snvoc:hasCreator ?creator.
    ?otherMessage snvoc:hasCreator ?creator;
                  snvoc:content ?messageContent.
} LIMIT 10
```

**Listing 4: SPARQL template for discover query 8.**

**Metrics**

We consider the following performance metrics in SolidBench:

- **Query execution time**: Time between sending the query to the engine, and obtaining the final result.

- **Query result arrival times**: For each result, time between sending the query, and obtaining that result.

- **HTTP requests**: For a query, the number of HTTP requests the engine issued during its execution.
- **Accuracy**: The F1-measure as a percentage indicating the correctness (precision) and completeness (recall) of query results with respect to the expected query results.

For measuring query execution and result arrival times, a warmup round with all instantiated query templates must take place first. To ensure that we take into account the volatile nature of the Web and the live traversal property of LTQP, the HTTP cache of the query engine is flushed before every query execution, while this cache can still be used within the span of a single query execution.

## 5  APPROACH

In this section, we introduce techniques for handling Solid's structural properties discussed in Section 3. Our goal is not to introduce additional components or structural properties to the Solid ecosystem, but instead, we use what the Solid ecosystem provides today, and investigate how to query over this as performant as possible. We start by discussing the preliminaries of the formalities we will introduce. Next, we discuss our pipeline-based link queue approach. Then, we discuss two novel discovery approaches for LTQP. Finally, we discuss their implementations.

### 5.1  Formal Preliminaries

This section summarizes the semantics of SPARQL query execution [57] and LTQP [14,31] that we build upon.

The infinite set of *RDF triples* is formalized as $\mathcal{T} = (\mathcal{I} \cup \mathcal{B}) \times \mathcal{I} \times (\mathcal{I} \cup \mathcal{B} \cup \mathcal{L})$, where $\mathcal{I}$, $\mathcal{B}$, and $\mathcal{L}$ respectively denote the disjoint, infinite sets of IRIs, blank nodes, and literals. Furthermore, $\mathcal{V}$ is the infinite set of all variables that is disjoint from $\mathcal{I}$, $\mathcal{B}$, and $\mathcal{L}$. A tuple $tp \in (\mathcal{V} \cup \mathcal{I}) \times (\mathcal{V} \cup \mathcal{I}) \times (\mathcal{V} \cup \mathcal{I} \cup \mathcal{L})$ is called a *triple pattern*. A finite set of these triple pattern is called a *basic graph pattern* (BGP). More complex SPARQL query operators exist, but since BGPs form the foundational building block of a SPARQL query, we only consider BGPs for the remainder of this work. The query results of a SPARQL query $P$ over a set of RDF triples $G$ are called *solution mappings*, which are denoted by $[[P]]_G$, consisting of partial mappings $\mu : \mathcal{V} \to (\mathcal{I} \cup \mathcal{B} \cup \mathcal{L})$. An RDF triple $t$ *matches* a triple pattern $tp$ if $\exists \mu : t = \mu[tp]$, where $\mu[tp]$ is the triple pattern that is obtained by replacing all variables from $\mu$ in $tp$.

Formally, the reachability approaches that were discussed in Section 2 define which links should be followed during link traversal, and are usually captured as *reachability criteria* [14]. However, this formalization is not expressive enough for only following specific URIs within only subject, predicate, or object in data triples. Therefore, we formalize new reachability criteria in this work as *source selectors* within the subweb specification formalization [31] that *is* expressive enough to capture this. Within this formalization, a source selector $\sigma$ is defined as $\sigma : \mathcal{W} \to 2^{\mathcal{I}}$, where $\mathcal{W}$ is a Web of Linked Data. The Web of Linked Data $\mathcal{W}$ is a tuple $\langle D, data, adoc \rangle$, where $D$ is a set of documents, $data$ a function from $D$ to $2^{\mathcal{T}}$ such that $data(d)$ is finite for each $d \in D$, and $adoc$ a partial dereferencing function from $\mathcal{U}$ to $D$.

Based on these definitions, we define the set of all Solid data vaults as $\Upsilon$, where each Solid data vault $v \in \Upsilon$ is defined as a set of triples, where $triples(v) \subseteq \mathcal{T}$. For a Solid data vault $v_{LDP}$ exposed through the LDP interface, the triples contained in such a Solid vault are captured in different documents $D_v \subseteq D$. Hereby, $triples(v_{LDP}) = \{t \mid \forall d \in D_v \wedge t \in data(d)\}$.

### 5.2  Pipeline-Based Link Queue

To execute a query, our approach builds upon the zero-knowledge query planning technique [25] to construct a logical query plan ahead of query execution. This resulting plan produces a tree of logical query operators representing the query execution order. To execute this plan, the logical operators are executed by specific physical operators. Our physical query execution builds upon the iterator-based pipeline approach [20], which is the most popular among LTQP implementations [58,42,59]. We consider the execution plan as a pipeline [60] of iterator-based physical operators, through which intermediary results flow through chained operators to produce results in a pull-based manner.

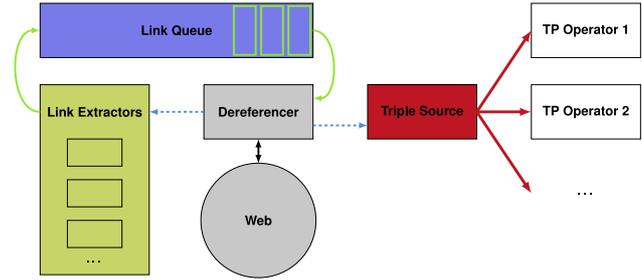

**Fig. 1: Link queue, dereferencer and link extractors feeding triples into a triple source, producing a stream of triples to tuple-producing operators in a pipelined query execution.**

Instead of letting operators trigger the dereferencing of URIs [20], we follow a link queue-based approach [32]. The architecture of this approach is visualized in Fig. 1. Concretely, we consider a continuously growing *triple source* as the basis of the pipeline tree, which is able to produce a (possibly infinite) stream of RDF triples. This triple source is fed triples originating from a loop consisting of the *link queue*, *dereferencer*, and a set of *link extractors*. The link queue accepts links from a set of link extraction components, which are invoked for every document that has been dereferenced by the dereferencer. The dereferenced documents containing triples are also sent to the continuously growing triple source. This link queue is initialized with a set of seed URIs, and the dereferencer continuously dereferences the URIs in the queue until it is empty. Since the link extractors are invoked after every dereference operation, this queue may virtually become infinitely long.

This link queue and link extractor approach is generic enough to implement other LTQP methods [20,14,30,31,32] for determining and prioritizing links that need to be followed. For example, one extractor may consider `rdfs:seeAlso` links, while another extractor may consider URIs of a triple that matches with a triple pattern from the query. Optionally, operators in the query pipeline may push links into the link queue, which enables context-based reachability semantics [30]. Link extractors only consider URIs as links, and ignore matching blank nodes and literals.

The triple source is connected to all tuple-producing SPARQL operators [57] in the leaves of the query plan, such as triple patterns and property path operators, into which a stream of triples is sent. The source indexes all triples, to ensure that an operator that is executed later in the execution process does not miss any triples.

### 5.3  Discovery Of Data Vault

In this section, we introduce a novel discovery approach for traversing over Solid data vaults as discussed in Section 3.

#### 5.3.1  Intuitive Description

To achieve link traversal within a vault, we assume that the WebID document is available as seed URI, or is discovered through some other reachability approach. As discussed in Section 3, the root of a vault can be discovered from a WebID document by dereferencing the object URI referred to by the `pim:storage` predicate. Next, all resources within this vault can be discovered by recursively following `ldp:contains` links from the root container.

We only consider triples for the `pim:storage` and `ldp:contains` predicates that have the current document URI as subject. If subjects contain fragment identifiers, we only consider them if the current document URI had this fragment identifier as well before it was dereferenced. For example, if a WebID with fragment identifier `#me` was discovered, then we only consider triples with the document URI + `#me` as subject.

### 5.3.2 Formal Description

We can formalize our discovery approach for the roots of data vaults as a following source selector starting from a given WebID with URI $i$ as $\sigma_{\text{SolidVault}}(W) = \{o \mid \langle i \text{ pim:storage } o \rangle \in data(adoc(i))\}$. Disjunctively coupled with this, we can formalize a source selector that can recursively traverse an LDP container as $\sigma_{\text{LdpContainer}}(W) = \{o \mid \forall s : \langle s \text{ ldp:contains } o \rangle \in data(adoc(s))\}$

### 5.4 Discovery Of Type Index

As discussed in Section 3, the type index enables resource discovery in a vault via RDF classes. In this section, we introduce a novel method that follows links in the type index, with an optional filter that only follows those links matching with a class in the query.

### 5.4.1 Intuitive Description

As before, we consider a WebID document as the starting point. From this document, we follow the `solid:publicTypeIndex` and `solid:privateTypeIndex` links. For each discovered type index, we consider all `solid:TypeRegistration` resources, and follow their `solid:instance` and `solid:instanceContainer` links.

As an optimization, we can also take into account the type information within the registrations of the type index, to only follow those links for classes that are of interest to the current query. Concretely, this involves considering the objects referred to by `solid:forClass` on each type registration. To know whether or not a class is relevant to the current query, we explicitly check for the occurrence of this class within the query as object within triples using the `rdf:type` predicate. For subjects in the query without `rdf:type` predicate, the matching class is unknown, which is why we consider all type registrations in this case.

### 5.4.2 Formal Description

To discover type indexes and follow links within them, we formalize the following source selector from a given WebID with URI $s$ when querying a BGP $B$:

$$\sigma_{\text{SolidTypeIndex}}(W) = \{o \mid \forall t, r, c : \phi(B, c)$$
$$\wedge \ (\langle s \text{ solid:publicTypeIndex } t \rangle \vee \langle s \text{ solid:privateTypeIndex } t \rangle)$$
$$\in data(adoc(s))$$
$$\wedge \ (\langle r \text{ rdf:type solid:TypeRegistration} \rangle$$
$$\wedge \langle r \text{ solid:forClass } c \rangle) \in data(adoc(t))$$
$$\wedge \ (\langle r \text{ solid:instance } o \rangle \vee \langle r \text{ solid:instanceContainer } o \rangle)$$
$$\in data(adoc(t))\}$$

Since `solid:instanceContainer` links to LDP containers, $\sigma_{\text{SolidTypeIndex}}$ should be disjunctively combined with $\sigma_{\text{LdpContainer}}$.

In this formalization, we consider $\phi(B, c)$ a filtering predicate function for determining which classes are considered within the type index. To consider *all* type registrations within the type index, we can implement $\phi(B, c)$ as a predicate always returning `true`. To only consider type registrations that match with a class mentioned in the query, we introduce the following filtering function:

$$\phi_{\text{QueryClass}}(B, c) = \begin{cases} \text{true} & \text{if } \exists tp \in B : \\ & \langle ?v \text{ rdf:type } c \rangle \text{ matches } tp \\ & \text{or if } \exists s : \langle s ? p ? o \rangle \in B \\ & \wedge \{o \mid \langle s \text{ rdf:type } o \in B \} = \emptyset \\ \text{false} & \text{else.} \end{cases}$$

### 5.5 Implementation

We have implemented our system using Comunica [61], which is a modular open-source SPARQL query engine framework. Concretely, we have implemented the pipeline-based link queue as a separate module, and we provide link extractors corresponding to the source selectors introduced in previous sections. Our implementation has full SPARQL 1.1 support, and consists of pipelined implementations of all monotonic SPARQL operators. This pipelined implementation is important for iterative tuple processing in a non-blocking manner, because the link queue and the resulting stream of triples may become infinitely long.

Our implementation focuses on the SPARQL query language, and does not make use of alternative LTQP-specific query languages or non-standard SPARQL language extensions such as LDQL [34] and SPARQL-LD [62] that incorporate link navigation paths into the query. As discussed in Section 3, different Solid apps or user preferences may lead to the storage of similar data at different locations within vaults. Hence, it is important that link navigation is *decoupled* from the query to keep queries reusable for different Solid users, as link paths to data may differ across different data vaults. Instead, our implementation uses LDP container traversal and the Solid type index to replace explicit navigation links.

To provide a stable reference implementation that can be used for the experiments in this work and future research, our implementation focuses on extensibility and reusability. We do this by implementing all logic in configurable modules that are extensively tested through integration and unit tests with 100% code coverage. Our implementation is available as open-source at https://github.com/comunica/comunica-feature-link-traversal.

Our implementation builds upon best practises in LTQP and lessons learned from other implementations [58] including, the use of client-side caching [41], the different reachability semantics [14], zero-knowledge query planning [25] applied to arbitrary join operations instead of only triple patterns in BGPs, and more [20]. Furthermore, our implementation allows users to explicitly pass seed URIs, but falls back to query-based seed URIs [58] if no seeds were provided. This fallback finds all URIs within the query, and adds them as seed URIs to the link queue.

Hence, this implementation meets the requirements of a query engine that can query over one or more Solid data vaults, as discussed in Section 3. This also includes the ability to perform authenticated to documents within vaults behind access control. To ensure that common HTTP errors that may occur during link traversal don't terminate the query execution process, we enable a default *lenient* mode, which ignores dereference responses with HTTP status code in ranges 400 and 500.

## 6 EVALUATION

In this section, we tackle the research question *"How well does link traversal query processing perform over decentralized environments with structural properties"*. Within this work, we apply our experiments to the structural properties of the decentralized environment provided by Solid, but findings may be generalizable to other decentralized environments. We provide an answer to this research question by simulating Solid data vaults using the benchmark introduced in Section 4 using the default configuration, and evaluating different approaches based on the implementation discussed in Section 5. We first introduce the design of our experiment, followed by presenting our experimental results, and a discussion of our results to answer our research question.

### 6.1 Experimental Design

We make use of a factorial experiment containing the following factors and values:

- **Vault discovery**: None, LDP, Type Index, Filtered Type Index, LDP and Type Index, LDP and Filtered Type Index
- **Reachability semantics**: cNone, cMatch, cAll
- **Fragmentation strategy**: Composite
- **Multiplication factor**: 1

The LDP strategy corresponds to the disjunction of the source selectors $\sigma_{\text{SolidVault}}$ and $\sigma_{\text{LdpContainer}}$, the Type Index to $\sigma_{\text{LdpContainer}}$ and $\sigma_{\text{SolidTypeIndex}}$ with $\phi(B, c)$ always returning true, and the Filtered Type Index to $\sigma_{\text{LdpContainer}}$ and $\sigma_{\text{SolidTypeIndex}}$ with $\phi_{\text{QueryClass}}$.

Furthermore, to measure the impact of the different fragmentation strategies that were discussed in Section 4, we compare them using the optimal method of vault discovery and reachability semantics (as determined later in the experiments):

- **Vault discovery**: LDP and Filtered Type Index
- **Reachability semantics**: cMatch

- › **Fragmentation strategy**: Separate, Single, Location, Time, Composite
- › **Multiplication factor**: 1, 5

Our experiments were performed on a 64-bit Ubuntu 14.04 machine with a 24-core 2.40 GHz CPU and 128 GB of RAM. The Solid vaults and query client were executed in isolated Docker containers on dedicated CPU cores with a simulated network. To foster reproducibility, the experimental setup, raw results, and processing scripts are available as open-source on https://github.com/comunica/Experiments-Solid-Link-Traversal. All queries were configured with a timeout of two minutes, and were executed three times to average metrics over. Each query template in the benchmark was instantiated five times, which resulted in 40 discover queries, 35 short queries, and 60 complex queries.

We were unable to compare our implementation to existing LTQP engines, because those systems (e.g. Lidaq [24]) would either require significant changes to work over Solid vaults, they depend on a non-standard usage of the SPARQL syntax (e.g. SPARQL-LD [62]), or insufficient documentation was present to make them work (e.g. SQUIN [58]), even after contacting the authors. Nevertheless, in order to ensure a fair and complete comparison, we have re-implemented the foundational LTQP algorithms (cNone, cMatch, cAll), and compare them against, and in combination with, our algorithms.

## 6.2 Experimental Results

In this section, we present results that offer insights into our research question. Table 2, Table 3, and Table 4 show the aggregated results for the different combinations of our setup for the discover, short, and complex queries of the benchmark, respectively. Furthermore, Table 5 shows the aggregated results of all discover queries over different fragmentation strategies with different post multiplication factors. Concretely, each table shows the average ($\bar{t}$) and median ($\tilde{t}$) execution times (ms), the average ($\bar{t}_1$) and median ($\tilde{t}_1$) time until first result (ms), average number of HTTP requests per query ($\overline{req}$), total number of results on average per query ($\sum ans$), average accuracy ($\overline{acc}$), and number of timeouts ($\sum to$) across all queries. The combinations with the highest accuracy value are marked in bold. The number of HTTP requests is counted across all query executions that did not time out within each combination. The timeout column represents the number of query templates that lead to a timeout for a given combination. The accuracy of each query execution is calculated as a percentage indicating the precision and recall of query results with respect to the expected query results.

| | $\bar{t}$ | $\tilde{t}$ | $\bar{t}_1$ | $\tilde{t}_1$ | $\overline{req}$ | $\sum ans$ | $\overline{acc}$ | $\sum to$ |
|---|---|---|---|---|---|---|---|---|
| cnone-base | 40 | 0 | N/A | N/A | 8 | 0.00 | 0.00% | 0 |
| cmatch-base | 1,791 | 0 | 22,946 | 24,439 | 1,275 | 0.00 | 0.00% | 1 |
| call-base | 128,320 | 127,021 | 28,448 | 10,554 | 0 | 0.63 | 3.13% | 8 |
| cnone-idx | 1,448 | 842 | 447 | 351 | 243 | 20.50 | 74.14% | 0 |
| **cmatch-idx** | 12,284 | 2,210 | 2,304 | 1,217 | 2,567 | 39.13 | 99.14% | 0 |
| call-idx | 124,197 | 124,811 | 48,223 | 9,778 | 18,022 | 3.13 | 17.40% | 7 |
| cnone-idx-filt | 1,429 | 755 | 435 | 311 | 230 | 20.50 | 74.14% | 0 |
| **cmatch-idx-filt** | 12,114 | 2,312 | 2,397 | 1,075 | 2,554 | 39.13 | 99.14% | 0 |
| call-idx-filt | 124,003 | 126,093 | 43,147 | 29,937 | 11,023 | 4.50 | 29.78% | 8 |
| cnone-ldp | 1,606 | 994 | 563 | 386 | 342 | 20.50 | 74.14% | 0 |
| cmatch-ldp | 13,463 | 2,288 | 3,660 | 1,057 | 3,625 | 37.88 | 86.64% | 1 |
| call-ldp | 123,712 | 123,479 | 37,083 | 13,733 | 0 | 2.00 | 16.25% | 8 |
| cnone-ldp-idx | 1,560 | 1,001 | 482 | 349 | 358 | 20.50 | 74.14% | 0 |
| **cmatch-ldp-idx** | 12,417 | 2,529 | 2,333 | 1,189 | 2,709 | 39.13 | 99.14% | 0 |
| call-ldp-idx | 127,768 | 125,103 | 67,577 | 13,472 | 466 | 2.38 | 16.63% | 7 |
| cnone-ldp-idx-filt | 1,552 | 1,006 | 425 | 331 | 357 | 20.50 | 74.14% | 0 |
| **cmatch-ldp-idx-filt** | 12,483 | 2,372 | 2,309 | 925 | 2,708 | 39.13 | 99.14% | 0 |
| call-ldp-idx-filt | 123,979 | 125,235 | 48,382 | 10,368 | 16,623 | 3.13 | 17.40% | 7 |

**Table 2: Aggregated results for the different combinations across all 8 discover queries.**

These results show that there are combinations of approaches that achieve a very high level of accuracy for discover queries, and an average level of accuracy for short queries. However, for complex queries, none of the combinations achieve an acceptable level of accuracy. Hence, we consider this last query category too complex for current link traversal approaches, and we do not consider them further in this article. Furthermore, increasing the number of posts within a vault has an increasing factor on the query execution times, but this factor varies with different fragmentation strategies. We will elaborate on these results in more detail hereafter.

| | $\bar{t}$ | $\tilde{t}$ | $\bar{t}_1$ | $\tilde{t}_1$ | $\overline{req}$ | $\sum ans$ | $\overline{acc}$ | $\sum to$ |
|---|---|---|---|---|---|---|---|---|
| cnone-base | 34,364 | 70 | 18 | 2 | 12 | 0.14 | 14.29% | 2 |
| cmatch-base | 47,700 | 987 | 121 | 92 | 592 | 0.43 | 42.86% | 3 |
| call-base | 126,794 | 125,609 | 1,547 | 787 | 0 | 0.00 | 0.00% | 7 |
| cnone-idx | 34,775 | 540 | 676 | 151 | 71 | 0.14 | 14.29% | 2 |
| cmatch-idx | 70,142 | 119,114 | 6,837 | 530 | 263 | 0.43 | 42.86% | 4 |
| call-idx | 109,943 | 123,227 | 14,290 | 19,345 | 0 | 0.00 | 0.00% | 7 |
| cnone-idx-filt | 34,804 | 534 | 527 | 110 | 71 | 0.14 | 14.29% | 2 |
| cmatch-idx-filt | 69,808 | 119,032 | 7,190 | 434 | 263 | 0.43 | 42.86% | 4 |
| call-idx-filt | 116,618 | 123,312 | 9,764 | 6,207 | 0 | 0.00 | 0.00% | 7 |
| cnone-ldp | 34,975 | 621 | 816 | 46 | 96 | 0.29 | 15.71% | 2 |
| **cmatch-ldp** | 70,026 | 119,586 | 6,234 | 636 | 291 | 0.57 | 44.29% | 4 |
| call-ldp | 127,550 | 126,587 | 717 | 483 | 0 | 0.00 | 0.00% | 7 |
| cnone-ldp-idx | 34,852 | 811 | 521 | 43 | 100 | 0.14 | 14.29% | 2 |
| cmatch-ldp-idx | 69,534 | 119,215 | 2,936 | 437 | 295 | 0.43 | 42.86% | 4 |
| call-ldp-idx | 110,217 | 122,525 | 8,841 | 6,114 | 0 | 0.00 | 0.00% | 7 |
| cnone-ldp-idx-filt | 34,830 | 742 | 402 | 83 | 100 | 0.14 | 14.29% | 2 |
| **cmatch-ldp-idx-filt** | 70,042 | 119,126 | 6,246 | 663 | 295 | 0.57 | 44.29% | 4 |
| call-ldp-idx-filt | 114,800 | 123,058 | 15,075 | 17,192 | 0 | 0.00 | 0.00% | 7 |

**Table 3: Aggregated results for the different combinations across all 7 short queries.**

| | $\bar{t}$ | $\tilde{t}$ | $\bar{t}_1$ | $\tilde{t}_1$ | $\overline{req}$ | $\sum ans$ | $\overline{acc}$ | $\sum to$ |
|---|---|---|---|---|---|---|---|---|
| cnone-base | 22,093 | 73 | N/A | N/A | 11 | 0.00 | 0.00% | 2 |
| cmatch-base | 81,008 | 119,633 | N/A | N/A | 999 | 0.00 | 0.00% | 9 |
| call-base | 125,270 | 124,949 | N/A | N/A | 0 | 0.00 | 0.00% | 12 |
| cnone-idx | 61,315 | 76,006 | N/A | N/A | 155 | 0.00 | 0.00% | 6 |
| cmatch-idx | 109,483 | 119,463 | N/A | N/A | 58 | 0.00 | 0.00% | 11 |
| call-idx | 124,760 | 123,723 | N/A | N/A | 0 | 0.00 | 0.00% | 12 |
| cnone-idx-filt | 61,370 | 74,490 | N/A | N/A | 155 | 0.00 | 0.00% | 6 |
| cmatch-idx-filt | 109,539 | 119,600 | N/A | N/A | 58 | 0.00 | 0.00% | 11 |
| call-idx-filt | 126,756 | 125,316 | N/A | N/A | 0 | 0.00 | 0.00% | 12 |
| cnone-ldp | 60,002 | 45,081 | N/A | N/A | 241 | 0.00 | 0.00% | 6 |
| cmatch-ldp | 117,589 | 123,023 | N/A | N/A | 55 | 0.00 | 0.00% | 11 |
| call-ldp | 127,184 | 126,186 | N/A | N/A | 0 | 0.00 | 0.00% | 12 |
| cnone-ldp-idx | 62,029 | 76,171 | N/A | N/A | 246 | 0.00 | 0.00% | 6 |
| cmatch-ldp-idx | 112,234 | 120,586 | N/A | N/A | 56 | 0.00 | 0.00% | 11 |
| call-ldp-idx | 125,116 | 124,494 | N/A | N/A | 0 | 0.00 | 0.00% | 12 |
| cnone-ldp-idx-filt | 59,081 | 43,802 | N/A | N/A | 246 | 0.00 | 0.00% | 6 |
| cmatch-ldp-idx-filt | 112,336 | 120,723 | N/A | N/A | 56 | 0.00 | 0.00% | 11 |
| call-ldp-idx-filt | 126,848 | 126,620 | N/A | N/A | 0 | 0.00 | 0.00% | 12 |

**Table 4: Aggregated results for the different combinations across all 12 complex queries.**

| | $\bar{t}$ | $\tilde{t}$ | $\bar{t}_1$ | $\tilde{t}_1$ | $\overline{req}$ | $\sum ans$ | $\sum to$ |
|---|---|---|---|---|---|---|---|
| Composite-1 | 15,207 | 2,804 | 5,154 | 921 | 2,653 | 39.13 | 0 |
| Composite-5 | 29,139 | 3,698 | 3,660 | 1,745 | 556 | 48.25 | 1 |
| Separate-1 | 3,310 | 2,297 | 2,092 | 594 | 2,681 | 39.13 | 0 |
| Separate-5 | 18,440 | 3,491 | 4,535 | 905 | 6,597 | 48.25 | 1 |
| Single-1 | 15,503 | 3,212 | 3,135 | 1,111 | 1,995 | 39.13 | 0 |
| Single-5 | 30,462 | 6,397 | 4,967 | 1,660 | 479 | 48.25 | 1 |
| Location-1 | 15,428 | 3,148 | 2,987 | 1,218 | 575 | 39.13 | 0 |
| Location-5 | 30,788 | 6,292 | 5,125 | 1,520 | 542 | 48.25 | 1 |
| Time-1 | 5,157 | 1,862 | 1,769 | 1,050 | 4,412 | 39.13 | 0 |
| Time-5 | 21,645 | 3,072 | 2,866 | 1,307 | 4,339 | 48.25 | 1 |

**Table 5: Aggregated results for the different fragmentation strategies over different post multiplication factors across all 8 discover queries.**

## 6.3 Discussion

### 6.3.1 Intra-Vault And Inter-Vault Data Discovery

The results above show that if we desire accurate results, that the combination of cMatch semantics together with at least one of the data vault discovery methods is required. This combination is needed because our workload contains queries that either target data within a single vault (e.g. D1), or data spanning multiple data vaults (e.g. D8). While the different data vault discovery methods are able to discover data *within* vaults, the reachability of cMatch is required to discover data *across* multiple vaults.

Due to this, cNone (follow no links) is an ineffective replacement for cMatch (follow links matching query) even when combined with discovery methods, because link traversal across multiple vaults will not take place, which will lead to too few query results. Concretely, for discover queries cNone can only achieve a accuracy of 74.14% for discover queries and 28.57% for short queries, compared to respectively 99.14% and 42.86% for cMatch. However, for those queries that target a single vault, cNone can be used instead of cMatch without a loss of accuracy, leading to a lower number of HTTP requests and lower query execution times.

Since cAll leads to all links being followed, including those followed by cMatch, it is theoretically a sufficient replacement for cMatch. However, our results show that too many links are being followed with cAll, which leads to timeouts for nearly all queries.

Our results show that solely using reachability semantics (cMatch or cAll) without a data discovery method is insufficient for discover queries, where a accuracy of only up to 3.13% can be achieved for discover queries. However, when looking at the short queries category, solely using reachability semantics appears to be sufficient, with the query execution time even being lower. This difference exists because the discover workload contains queries that discover data related to a certain person or resource, while the short queries target only details of specific resources. Discover queries therefore depend on an overview of the vault, while short queries only depend on specific links between resources within a vault. The remainder of this discussion only focuses on discover queries, since these achieve the highest level of accuracy. As such, the short and complex queries highlight opportunities for improvement in future work.

### 6.3.2 Type Index And LDP Discovery Perform Similarly

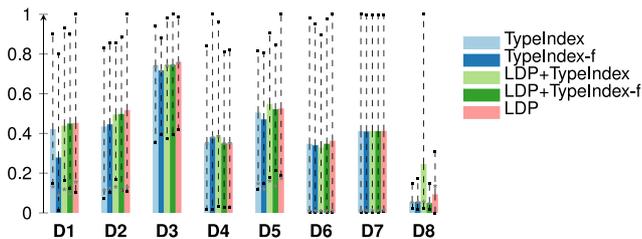

**Fig. 2: Relative execution times for discover queries with different discovery methods under cMatch. Bars indicate average execution time, whiskers indicate the maxima and minima, and stars indicate average time until first result.**

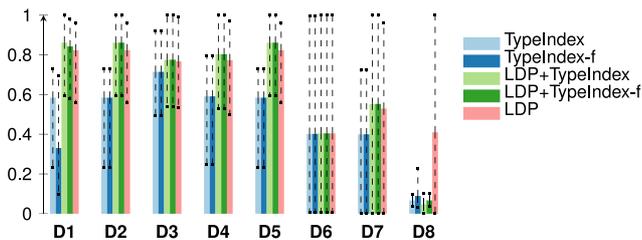

**Fig. 3: Relative number of HTTP requests for discover queries with different discovery methods under cMatch. Bars indicate average execution time, whiskers indicate the maxima and minima.**

When comparing the number of HTTP requests and query execution times for different data vault discovery approaches under cMatch in Table 2, we can observe that using the type index leads to fewer HTTP requests and faster query execution compared to LDP-based discovery on average. To explain this behaviour in more detail, Fig. 2 shows the average query execution times of each discover query separately, for the different combinations of data vault discovery approaches. To simplify comparability, the execution times within this figure are relative to the maximum query execution time per query [32]. Furthermore, Fig. 3 shows the average number of HTTP requests for each of those discover queries, which are also made relative to the maximum number of requests per query for better comparability. Fig. 4, Fig. 5, and Fig. 6 contain more detailed query result arrival times for several of these queries using diefficiency plots [63]. Finally, Table 6 shows an overview of the number of queries where each approach achieves the lowest execution time per query.

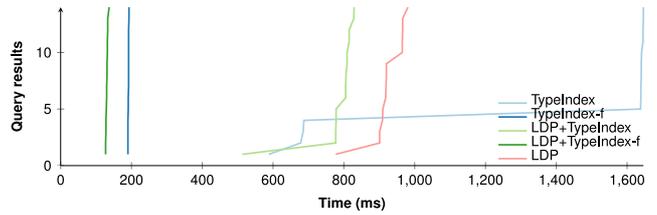

**Fig. 4: Query result arrival times for D1 with different combinations of data vault discovery with cMatch.**

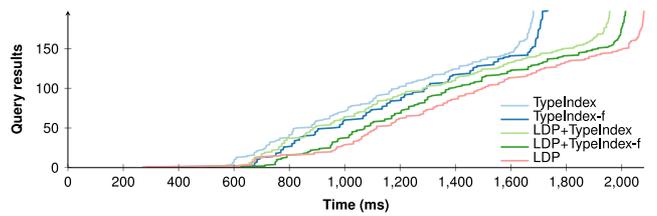

**Fig. 5: Query result arrival times for D2 with different combinations of data vault discovery with cMatch.**

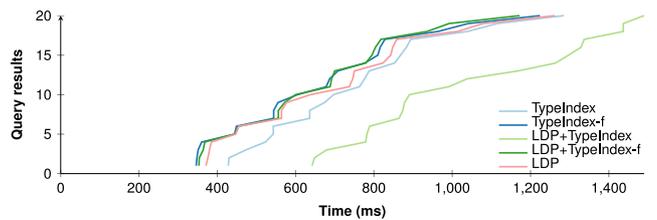

**Fig. 6: Query result arrival times for D5 with different combinations of data vault discovery with cMatch.**

| | TypeIndex | TypeIndex-f | LDP | LDP + TypeIndex | LDP + TypeIndex-f |
|---|---|---|---|---|---|
| Wins | 5 | 5 | 5 | 5 | 15 |

**Table 6: The number of queries each approach achieves the lowest query execution time for across all cMatch-based approaches over all 8 discover queries with 5 instantiations. A win for a certain approach is only considered if the results are accurate for this query. Five queries are missing due to no approaches achieving accurate results.**

While Fig. 2 shows that for all queries using just the type index is slightly faster or comparable to just LDP-based discovery, this difference has no statistical significance ($p = 0.40$). However, Fig. 3 shows that the number of HTTP requests with the type index is always significantly lower than via LDP ($p = 0.01$).

When the filter-enabled type index approach is used, five queries (D1, D3, D5, D6, D7) are made even faster compared to the non-filtered type index approach. This is because those queries target a possibly empty subset of the type index entries, which means that a significant range of links can be pruned out, which leads to a major reduction in the number of HTTP requests, which is a main bottleneck in link traversal. For the other queries, the filter-enabled approach becomes slightly slower than (D2, D4) or is comparable to (D8) the non-filtered type index approach. For those queries, the processing overhead of type index filtering becomes too high compared to its potential benefit. Statistically, this difference has no significance in terms of execution time ($p = 0.69$) and number of HTTP requests ($p = 0.68$).

These results show that using the type index together with LDP-based discovery is not *significantly* better than the other approach-es ($p = 0.71$), which is primarily caused by the statistically significantly higher number of HTTP requests ($p = 0.02$) required for traversing both the type index and nested LDP containers. Query D8 and Table 6 does however show that this combination deserves further investigation, because this query has a result limit that leads to a prioritization of links via the type index, leading to earlier query termination with fewer requests.

Fig. 4, Fig. 5, and Fig. 6 show the query result arrival times for D1, D2, and D5 respectively with different combinations of data vault discovery with cMatch. Since D1 specifically queries for resources of the type Post, it can very selectively make use of the Post entry within the type index, which makes the filtered type index approach faster than the non-filtered approach. D2 targets both resources of type Comment and Post, which means that it has to make use of both entries within the type index, which causes the performance difference between the filtered and non-filtered type index approach to be negligeable. D5 is a query that does not specifically target resources of certain types. This means that the type index leads to no significant performance benefit if no specific types are targeted in the query.

In general, these results hint that the LDP-based approach combined with filtered type index approach performs better than the other approaches. However, due to the minimal difference in terms of execution time, the performance of all approaches can be considered equivalent.

### 6.3.3 Zero-Knowledge Query Planning Is Ineffective

While it may seem obvious to assume that higher query execution times are caused by a higher number of links that need to be dereferenced, we observe only a weak correlation ($\rho = 0.32$) of this within the cMatch-based discovery approaches discussed before. As such, the main bottleneck in this case appears not primarily to be the number of links to traverse. Instead, our analysis suggests that query plan efficiency is the primary influencer of execution times.

To empirically prove this finding, we compare the execution times of our default integrated query execution approach (cMatch with filtered type index discovery) with a two-phase query execution approach that we implemented in the same query engine. Instead of following links during query execution as in the integrated approach, the two-phase approach first follows links to index all discovered triples, and processes the query in the traditional *optimize-then-execute* manner. This two-phase approach is based on an oracle that provides all query-relevant links, which we determined by analyzing the request logs during the execution of the integrated approach. Therefore, this two-phase approach is merely a theoretical case, which delays time until first results due to prior indexing, and which may not always be achievable in practise due to infinitely growing link queues for some queries. The results of this experiment are shown in Table 7.

| Query | Integrated | Two-phase | HTTP Requests |
|---|---|---|---|
| D1 | 1,077.58 | 403.54 | 222 |
| D2 | 1,020.67 | 567.57 | 223 |
| D3 | 1,193.01 | 821.23 | 429 |
| D4 | 3,266.62 | 505.00 | 228 |
| D5 | 522.23 | 387.24 | 223 |
| D6 | 710.16 | 289.72 | 122 |
| D7 | 626.96 | 340.54 | 122 |
| D8 | 2,037.85 | 1,654.02 | 420 |

**Table 7: Integrated and two-phase execution times (ms) of discover queries, with number of HTTP requests per query.**

These results show that the two-phase approach is on average two times faster for all queries compared to the integrated approach, even when taking into account time for dereferencing. The reason for this is that the two-phase approach is able to perform traditional query planning [64,65], since it has access to an indexed triple store with planning-relevant information such as cardinality estimates. Since the integrated approach finds new triples *during* query execution, it is unable to use this information for traditional query planning. Instead, our integrated approach makes use of the zero-knowledge query planning technique [25] that makes use of heuristics to plan the query before execution.

Since the only difference between the implementations of the integrated and two-phase approach is in how they plan the query, we can derive the query plan of the integrated approach is very ineffective. As such, there is clear need for better query planning during integrated execution, and the two-phase approach shows that performance could become more than two times better.

Zero-knowledge query planning [25] is ineffective in our experiments because it has been designed under the assumptions of Linked Open Data, while it does not match with the structural assumptions of specific decentralized environments such as Solid. For example, one of the heuristics within this planner deprioritizes triple patterns with vocabulary terms, such as rdf:type, since they are usually the least selective. However, when a Solid type index is present, such vocabulary terms may instead become *very selective*, which means that those would benefit from prioritization. As such, there is a need for alternative query planners that consider the structural assumptions within specific decentralized environments.

### 6.3.4 Vault Size And Fragmentation Impact Performance

The results in Table 5 show that different fragmentation strategies with different multiplication factors for vault sizes can impact both execution times and the number of HTTP requests. To enable better comparisons, Fig. 7 and Fig. 8 respectively show the average query execution times and number of HTTP requests of each discover query separately. Furthermore, Fig. 9 and Fig. 10 contain die efficiency plots for some of these queries.

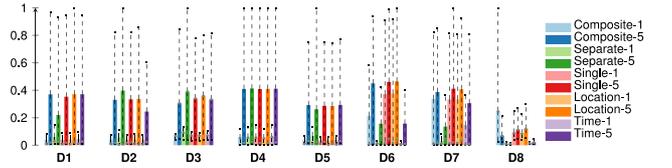

**Fig. 7: Relative execution times for discover queries with different fragmentation strategies and multiplication factors under cMatch. Bars indicate average execution time, whiskers indicate the maxima and minima, and stars indicate average time until first result.**

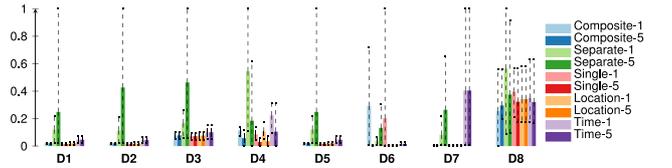

**Fig. 8: Relative number of HTTP requests for discover queries with different fragmentation and multiplication factors strategies under cMatch. Bars indicate average execution time, whiskers indicate the maxima and minima.**

These findings show that fragmenting data in different ways has a significant impact on the number of HTTP requests ($p < 0.01$). However, this does not translate into a significant difference in query execution times ($p = 0.72$). When we increase the amount of data within each pod, we *do* see a a significant difference in both execution times ($p = 0.014$) *and* number of HTTP requests ($p < 0.01$). In general, there is no significant correlation between the number of HTTP requests and the execution times ($p = 0.18$).

Since the *separate* fragmentation strategy produces separate files per post, it to be expected that this strategy results in the highest number of HTTP requests most queries, and is aggreveted when the vault size increases. This does not translate into it always leading to the highest execution times. However, as can be seen in Table 5, this strategy leads to the lowest times until first result.

This behaviour can be confirmed when inspecting the diefficiency plots in Fig. 9, Fig. 10. These show that the *separate* strategy usually leads to very early results, but later results come in relatively slowly, which causes other strategies that emit their first result later, to still achieve a lower total execution time. This is because this strategy leads to many very small files, each of which can be fetched and processed very efficiently. But due to their high number, fetching many of them incurs an overhead in terms of HTTP requests.

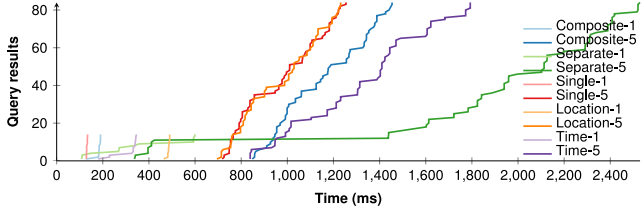

**Fig. 9: Query result arrival times for D1 with different fragmentation strategies and multiplication factors.**

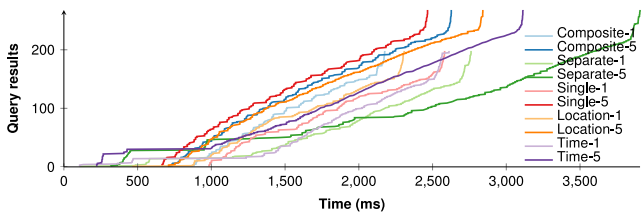

**Fig. 10: Query result arrival times for D2 with different fragmentation strategies and multiplication factors.**

While these results may indicate that some fragmentation strategies are more favorable than others, the strategy used within a user's vault is usually not something the query engine can influence. In decentralized environments such as Solid, users are in full control over their data, which means that query engines should not enforce their needs for efficient execution onto the user. The engine may however handle certain strategies more efficient than others, and could inform the user if non-optimal fragmentation strategies are encountered in vaults. In general, query engines should not make assumptions about the fragmentation strategy in vaults. Instead, engines must apply live exploration of vaults to handle heterogeneous fragmentations.

## 7 CONCLUSIONS

User-oriented decentralized applications require results in the order of seconds or less to avoid losing the user's attention [51]. Our work has shown that Link Traversal Query Processing is able to achieve such timings, especially as it is able to produce results in an iterative manner, with first results mostly being produced in less than a second. As such, LTQP with the algorithms introduced in this work are effective for querying over decentralized environments with specific structural properties, but there are open research opportunities for optimizing more complex queries as provided by our benchmark. We have shown this by applying LTQP to simulated Solid environments, for which we have introduced specific algorithms for capturing these structural properties.

Up until now, LTQP has always been applied to querying Linked Open Data on the Web. In that context, it was assumed that *"we should never expect complete results"* [20], because of the openness and distributed nature of the Web. However, when applying LTQP to specific decentralized environments such as Solid, this limitation does not hold anymore, since there are additional assumptions that we can make use of during query execution. For instance, the ability to close the world around Solid vaults, and the data discovery techniques that Solid vaults provide, create opportunities for query execution that allow us to guarantee complete results. While we have investigated the specific case of querying Solid vaults, these concepts may be generalizable to other decen-tralization efforts, such as Mastodon [3]. This is possible, because our approach solely relies on the structural properties provided by standards such as the Linked Data Platform [15] and the Type Index [16], which can be used outside of the Solid ecosystem.

Due to the possibility of incomplete results, LTQP research over Linked Open Data is moving into the direction of finding query-relevant documents as early as possible [32]. In the context of Solid, we have shown that finding all query-relevant documents is not the main bottleneck during query execution anymore. Instead, the *effectiveness of the query plan* is has become the new bottleneck. While finding query-relevant documents is still relevant for specific decentralized environments, we show the need for more research towards better query planning techniques. Since LTQP leads to data being discovered during query execution, adaptive query planning [66] techniques are highly promising. So far, these techniques have only seem limited adoption within LTQP [32] and SPARQL query processing [67, 68, 69]. For example, we see many opportunities in decomposing parts of the query plan over Solid type index entries upon index discovery, which requires a type index-aware adaptive query plan optimizer.

Our findings about discovery approaches such as the Solid type index have a great potential for improving query performance. However, we have only scratched the surface of what is possible in this work. On the one hand, usage of this type index can be further improved. While we only made us of this index for explicit `rdf:type` occurrences in the query, this could be extended to also consider implicit type knowledge via inferencing [70]. Furthermore, query containment techniques [71] could be investigated for determining which parts of the query match with index entries, which could also lead to better a combination of type index and LDP-based discovery. On the other hand, alternative structural properties could offer more expressivity, such as characteristics sets [72] and other summarization techniques [24]. For example, the Shape Trees [73] specification offers a similar index for Solid vaults, but instead makes use of nested ShEx [74] shape definitions for expressing data locations, which could be used for both data discovery and query optimization [75]. Next to that, the incorporation of more expressive Linked Data Fragments interfaces [76, 77, 78, 79, 80, 81] in certain Solid vaults could introduce interesting trade-offs in terms of server and client query execution effort. However, since different vaults could expose different types of interfaces, which are only discovered on the fly during query execution, novel adaptive query execution algorithms for heterogeneous interfaces [82, 83, 84] will be required. Finally, our work only considers read queries, while similar open questions remain for write-queries when updating data within decentralized environments.

This work provides an answer to the increasing need of querying techniques across decentralized environments, and uncovers the next steps that are needed for resolving current limitations. As such, this brings us one step closer towards a decentralized Web where users can be in full control over their data.

## ACKNOWLEDGEMENTS

The described research activities were funded by Ghent University and imec. Ruben Taelman is a postdoctoral fellow of the Research Foundation – Flanders (FWO) (1274521N).

## REFERENCES

[1] Berners-Lee, T.J.: Information management: A proposal. (1989).

[2] Mansour, E., Sambra, A.V., Hawke, S., Zereba, M., Capadisli, S., Ghanem, A., Aboulnaga, A., Berners-Lee, T.: A Demonstration of the Solid platform for Social Web Applications. In: Proceedings of the 25th International Conference Companion on World Wide Web. pp. 223–226. International World Wide Web Conferences Steering Committee (2016).

[3] Zignani, M., Gaito, S., Rossi, G.P.: Follow the Mastodon: Structure and Evolution of a Decentralized Online Social Network. In: Twelfth International AAAI Conference on Web and Social Media (2018).

[4] Kuhn, T., Taelman, R., Emonet, V., Antonatos, H., Soiland-Reyes, S., Dumontier, M.: Semantic micro-contributions with decentralized nanopublication services. PeerJ Computer Science. (2021). doi:10.7717/peerj-cs.387

[5] Hogan, A., Blomqvist, E., Cochez, M., d'Amato, C., Melo, G.de, Gutierrez, C., Kirrane, S., Gayo, J.E.L., Navigli, R., Neumaier, S., others: Knowledge graphs. Synthesis Lectures on Data, Semantics, and Knowledge. 12, 1–257 (2021).

[6] Berners-Lee, T.: Linked Data. https://www.w3.org/DesignIssues/LinkedData.html (2009).

[7] Cyganiak, R., Wood, D., Lanthaler, M.: RDF 1.1: Concepts and Abstract Syntax. W3C, https://www.w3.org/TR/2014/REC-rdf11-concepts-20140225/ (2014).

[8] Feigenbaum, L., Todd Williams, G., Grant Clark, K., Torres, E.: SPARQL 1.1 Protocol. W3C, https://www.w3.org/TR/2013/REC-sparql11-protocol-20130321/ (2013).

[9] Schwarte, A., Haase, P., Hose, K., Schenkel, R., Schmidt, M.: Fedx: Optimization Techniques for Federated Query Processing on Linked Data. In: International semantic web conference. pp. 601–616. Springer (2011).

[10] Verborgh, R., Vander Sande, M., Hartig, O., Van Herwegen, J., De Vocht, L., De Meester, B., Haesendonck, G., Colpaert, P.: Triple Pattern Fragments: a low-cost Knowledge Graph Interface for the Web. Journal of Web Semantics. 37, 184–206 (2016).

[11] Saleem, M., Ngomo, A.-C.N.: Hibiscus: Hypergraph-based Source Selection for SPARQL Endpoint Federation. In: European semantic web conference. pp. 176–191. Springer (2014).

[12] Görlitz, O., Staab, S.: Splendid: SPARQL Endpoint Federation Exploiting Void Descriptions. In: Proceedings of the Second International Conference on Consuming Linked Data-Volume 782. pp. 13–24. CEUR-WS. org (2011).

[13] Hartig, O.: An Overview on Execution Strategies for Linked Data Queries. Datenbank-Spektrum. 13, 89–99 (2013).

[14] Hartig, O., Freytag, J.-C.: Foundations of Traversal based Query Execution over Linked Data. In: Proceedings of the 23rd ACM conference on Hypertext and social media. pp. 43–52. ACM (2012).

[15] Speicher, S., Arwe, J., Malhotra, A.: Linked Data Platform 1.0. W3C, https://www.w3.org/TR/ldp/ (2015).

[16] Turdean, T.: Type Indexes. Solid, https://solid.github.io/type-indexes/ (2022).

[17] Prud'hommeaux, E., Carothers, G., Machina, L.: JSON-LD 1.1 Processing Algorithms and API. W3C, https://www.w3.org/TR/turtle/ (2014).

[18] Kellogg, G., Longley, D., Champin, P.-A.: JSON-LD 1.1 Processing Algorithms and API. W3C, https://www.w3.org/TR/json-ld/ (2020).

[19] Harris, S., Seaborne, A., Prud'hommeaux, E.: SPARQL 1.1 Query Language. W3C, https://www.w3.org/TR/2013/REC-sparql11-query-20130321/ (2013).

[20] Hartig, O.: SPARQL for a Web of Linked Data: Semantics and computability. In: Extended Semantic Web Conference. pp. 8–23. Springer (2012).

[21] Taelman, R., Verborgh, R.: A Prospective Analysis of Security Vulnerabilities within Link Traversal-Based Query Processing. In: Proceedings of the 6th International Workshop on Storing, Querying and Benchmarking Knowledge Graphs (2022).

[22] Hartig, O., Hose, K., Sequeda, J.: Linked Data Management. In: Sakr, S. and Zomaya, A. (eds.) Encyclopedia of Big Data Technologies. Springer, Germany (2019). doi:10.1007/978-3-319-63962-8_76-1

[23] Harth, A., Hose, K., Karnstedt, M., Polleres, A., Sattler, K.-U., Umbrich, J.: Data summaries for on-demand queries over linked data. In: Proceedings of the 19th international conference on World wide web. pp. 411–420 (2010).

[24] Umbrich, J., Hose, K., Karnstedt, M., Harth, A., Polleres, A.: Comparing data summaries for processing live queries over linked data. World Wide Web. 14, 495–544 (2011).

[25] Hartig, O.: Zero-knowledge query planning for an iterator implementation of link traversal based query execution. In: Extended Semantic Web Conference. pp. 154–169. Springer (2011).

[26] Mendelzon, A.O., Mihaila, G.A., Milo, T.: Querying the world wide web. In: Fourth International Conference on Parallel and Distributed Information Systems. pp. 80–91. IEEE (1996).

[27] Konopnicki, D., Shmueli, O.: Information gathering in the World-Wide Web: the W3QL query language and the W3QS system. ACM Transactions on Database Systems (TODS). 23, 369–410 (1998).

[28] Chakrabarti, S., Van den Berg, M., Dom, B.: Focused crawling: a new approach to topic-specific Web resource discovery. Computer networks. 31, 1623–1640 (1999).

[29] Batsakis, S., Petrakis, E.G.M., Milios, E.: Improving the performance of focused web crawlers. Data & Knowledge Engineering. 68, 1001–1013 (2009).

[30] Hartig, O., Pirrò, G.: SPARQL with Property Paths on the Web. Semantic Web. 8, 773–795 (2017).

[31] Bogaerts, B., Ketsman, B., Zeboudj, Y., Aamer, H., Taelman, R., Verborgh, R.: Link Traversal with Distributed Subweb Specifications. In: Rules and Reasoning: 5th International Joint Conference, RuleML+RR 2021, Leuven, Belgium, September 8 – September 15, 2021, Proceedings (2021).

[32]

[33] Schaffert, S., Bauer, C., Kurz, T., Dorschel, F., Glachs, D., Fernandez, M.: The linked media framework: Integrating and interlinking enterprise media content and data. In: Proceedings of the 8th International Conference on Semantic Systems. pp. 25–32 (2012).

[34] Hartig, O., Pérez, J.: LDQL: A query language for the web of linked data. Journal of Web Semantics. 41, 9–29 (2016).

[35] Fionda, V., Pirrò, G., Gutierrez, C.: NautiLOD: A formal language for the web of data graph. ACM Transactions on the Web (TWEB). 9, 1–43 (2015).

[36] Taelman, R., Steyskal, S., Kirrane, S.: Towards Querying in Decentralized Environments with Privacy-Preserving Aggregation. In: Proceedings of the 4th Workshop on Storing, Querying, and Benchmarking the Web of Data (2020).

[37] Umbrich, J., Hogan, A., Polleres, A., Decker, S.: Link Traversal Querying for a diverse Web of Data. Semantic Web. 6, 585–624 (2015).

[38] Schmidt, M., Görlitz, O., Haase, P., Ladwig, G., Schwarte, A., Tran, T.: Fedbench: A benchmark suite for federated semantic data query processing. In: International Semantic Web Conference. pp. 585–600. Springer (2011).

[39] Morsey, M., Lehmann, J., Auer, S., Ngonga Ngomo, A.-C.: DBpedia SPARQL benchmark–performance assessment with real queries on real data. In: International semantic web conference. pp. 454–469. Springer (2011).

[40] Görlitz, O., Thimm, M., Staab, S.: SPLODGE: Systematic Generation of SPARQL Benchmark Queries for Linked Open Data. In: ISWC (1). pp. 116–132 (2012).

[41] Hartig, O.: How caching improves efficiency and result completeness for querying linked data. In: LDOW (2011).

[42] Ladwig, G., Tran, T.: SIHJoin: Querying remote and local linked data. In: Extended Semantic Web Conference. pp. 139–153. Springer (2011).

[43] Capadisli, S., Berners-Lee, T., Verborgh, R., Kjernsmo, K.: Solid Protocol. Solid, https://solidproject.org/TR/protocol (2020).

[44] Van Herwegen, J., Verborgh, R., Taelman, R., Bosquet, M.: Community Solid Server. https://github.com/CommunitySolidServer/CommunitySolidServer (2022).

[45] Inrupt: PodSpaces. https://docs.inrupt.com/pod-spaces/ (2022).

[46] Flanders, D.: The Flemish Data Utility Company. https://www.vlaanderen.be/digitaal-vlaanderen/het-vlaams-datanutsbedrijf/the-flemish-data-utility-company (2022).

[47] Capadisli, S.: Web Access Control. Solid, https://solid.github.io/web-access-control-spec/ (2022).

[48] Bosquet, M.: Access Control Policy (ACP). Solid, https://solid.github.io/authorization-panel/acp-specification/ (2022).

[49] Coburn, A., Pavlik, elf, Zagidulin, D.: Solid-OIDC. Solid, https://solid.github.io/solid-oidc/ (2022).

[50] Capadisli, S., Berners-Lee, T.: Solid WebID Profile. Solid, https://solid.github.io/webid-profile/ (2022).

[51] Nielsen, J.: Response times: the three important limits. Usability Engineering (1993).

[52] Angles, R., Boncz, P., Larriba-Pey, J., Fundulaki, I., Neumann, T., Erling, O., Neubauer, P., Martinez-Bazan, N., Kotsev, V., Toma, I.: The linked data benchmark council: a graph and RDF industry benchmarking effort. ACM SIGMOD Record. 43, 27–31 (2014).

[53] Cheng, S., Hartig, O.: LinGBM: A Performance Benchmark for Approaches to Build GraphQL Servers. In: Proceedings of the 23rd International Conference on Web Information Systems Engineering (WISE 2022)

[54] Erling, O., Averbuch, A., Larriba-Pey, J., Chafi, H., Gubichev, A., Prat, A., Pham, M.-D., Boncz, P.: The LDBC social network benchmark: Interactive workload. In: Proceedings of the 2015 ACM SIGMOD International Conference on Management of Data. pp. 619–630 (2015).

[55] Angles, R., Antal, J.B., Averbuch, A., Boncz, P., Erling, O., Gubichev, A., Haprian, V., Kaufmann, M., Pey Josep Lluís Larriba, Martínez Norbert, others: The LDBC social network benchmark. arXiv preprint arXiv:2001.02299. (2020).

[56] Taelman, R., Van Herwegen, J., Vander Sande, M., Verborgh, R.: Components.js: Semantic Dependency Injection. Semantic Web Journal. (2022).

[57] Pérez, J., Arenas, M., Gutierrez, C.: Semantics and complexity of SPARQL. ACM Transactions on Database Systems (TODS). 34, 1–45 (2009).

[58] Hartig, O.: SQUIN: a traversal based query execution system for the web of linked data. In: Proceedings of the 2013 ACM SIGMOD International Conference on Management of Data. pp. 1081–1084 (2013).

[59] Miranker, D.P., Depena, R.K., Jung, H., Sequeda, J.F., Reyna, C.: Diamond: A SPARQL query engine, for linked data based on the

Hartig, O., Özsu, M.T.: Walking without a Map: Optimizing Response Times of Traversal-based Linked Data Queries (extended version). arXiv preprint arXiv:1607.01046. (2016).

Rete match. In: Proc. of the Workshop on Artificial Intelligence meet: Web of Data (AImWD) (2012).

60. Wilschut, A.N., Apers, P.M.G.: Pipelining in query execution. In: Proceedings. PARBASE-90: International Conference on Databases, Parallel Architectures, and Their Applications. p. 562. IEEE (1990).

61. Taelman, R., Van Herwegen, J., Vander Sande, M., Verborgh, R.: Comunica: a Modular SPARQL Query Engine for the Web. In: Proceedings of the 17th International Semantic Web Conference (2018).

62. Fafalios, P., Yannakis, T., Tzitzikas, Y.: Querying the Web of Data with SPARQL-LD. In: Research and Advanced Technology for Digital Libraries: 20th International Conference on Theory and Practice of Digital Libraries, TPDL 2016, Hannover, Germany, September 5–9, 2016, Proceedings 20. pp. 175–187. Springer (2016).

63. Acosta, M., Vidal, M.-E., Sure-Vetter, Y.: Diefficiency metrics: measuring the continuous efficiency of query processing approaches. In: International Semantic Web Conference. pp. 3–19. Springer (2017).

64. Schmidt, M., Meier, M., Lausen, G.: Foundations of SPARQL query optimization. In: Proceedings of the 13th International Conference on Database Theory. pp. 4–33. ACM (2010).

65. Stocker, M., Seaborne, A., Bernstein, A., Kiefer, C., Reynolds, D.: SPARQL Basic Graph Pattern Optimization using Selectivity Estimation. In: Proceedings of the 17th international conference on World Wide Web. pp. 595–604. ACM (2008).

66. Deshpande, A., Ives, Z., Raman, V.: Adaptive query processing. Foundations and Trends\textregistered in Databases. 1, 1–140 (2007).

67. Acosta, M., Vidal, M.-E., Lampo, T., Castillo, J., Ruckhaus, E.: ANAPSID: an adaptive query processing engine for SPARQL endpoints. In: International Semantic Web Conference. pp. 18–34. Springer (2011).

68. Acosta, M., Vidal, M.-E.: Networks of linked data eddies: An adaptive web query processing engine for RDF data. In: International Semantic Web Conference. pp. 111–127. Springer (2015).

69. Heling, L., Acosta, M.: Robust query processing for linked data fragments. Semantic Web. 1–35 (2022).

70. Kifer, M.: Rule interchange format: The framework. In: International Conference on Web Reasoning and Rule Systems. pp. 1–11. Springer (2008).

71. Wudage Chekol, M., Euzenat, J., Genevès, P., Layaïda Nabil: Evaluating and benchmarking SPARQL query containment solvers. In: International Semantic Web Conference. pp. 408–423. Springer (2013).

72. Neumann, T., Moerkotte, G.: Characteristic sets: Accurate cardinality estimation for RDF queries with multiple joins. In: 2011 IEEE 27th International Conference on Data Engineering. pp. 984–994. IEEE (2011).

73. Prud'hommeaux, E., Bingham, J.: Shape Trees Specification. W3C, https://shapetrees.org/TR/specification/ (2021).

74. Prud'hommeaux, E., Boneva, I., Emilio Labra Gayo, J., Kellogg, G.: Shape Expressions Language 2.1. W3C, https://shex.io/shex-semantics/ (2019).

75. Rabbani, K., Lissandrini, M., Hose, K.: Optimizing SPARQL queries using Shape Statistics. EDBT. (2021).

76. Verborgh, R., Vander Sande, M., Hartig, O., Van Herwegen, J., De Vocht, L., De Meester, B., Haesendonck, G., Colpaert, P.: Triple Pattern Fragments: a low-cost Knowledge Graph Interface for the Web. Journal of Web Semantics. 37, 184–206 (2016).

77. Azzam, A., Fernández, J.D., Acosta, M., Beno, M., Polleres, A.: SMART-KG: hybrid shipping for SPARQL querying on the web. In: Proceedings of The Web Conference 2020. pp. 984–994 (2020).

78. Minier, T., Skaf-Molli, H., Molli, P.: SaGe: Web preemption for public SPARQL query services. In: The World Wide Web Conference. pp. 1268–1278 (2019).

79. Azzam, A., Aebeloe, C., Montoya, G., Keles, I., Polleres, A., Hose, K.: WiseKG: Balanced access to web knowledge graphs. In: Proceedings of the Web Conference 2021. pp. 1422–1434 (2021).

80. Aebeloe, C., Keles, I., Montoya, G., Hose, K.: Star Pattern Fragments: Accessing Knowledge Graphs through Star Patterns. arXiv preprint arXiv:2002.09172. (2020).

81. Hartig, O., Buil-Aranda, C.: Bindings-restricted triple pattern fragments. In: OTM Confederated International Conferences" On the Move to Meaningful Internet Systems". pp. 762–779. Springer (2016).

82. Heling, L., Acosta, M.: Federated SPARQL Query Processing over Heterogeneous Linked Data Fragments. In: Proceedings of the ACM Web Conference 2022. pp. 1047–1057 (2022).

83. Cheng, S., Hartig, O.: FedQPL: A Language for Logical Query Plans over Heterogeneous Federations of RDF Data Sources. In: Proceedings of the 22nd International Conference on Information Integration and Web-based Applications & Services. pp. 436–445 (2020).

84. Montoya, G., Aebeloe, C., Hose, K.: Towards efficient query processing over heterogeneous RDF interfaces. In: 2nd Workshop on Decentralizing the Semantic Web, DeSemWeb 2018. CEUR Workshop Proceedings (2018).